\documentstyle[12pt]{article}

\title{Computational Geometry Column 33}
\author{Joseph O'Rourke%
\thanks{
	Department\ of\ Computer\ Science,
	Smith College,
	Northampton, MA 01063,
	USA.
	{\tt orourke@cs.smith.edu}.
	}
}
\date{}
\begin{document}
\bibliographystyle{alpha}
\maketitle
\pagestyle{empty}
\thispagestyle{empty}

\begin{abstract}
Several recent SIGGRAPH papers on surface simplification are described.
\end{abstract}

The stringent demands of real-time graphics have engendered a need
for simplification of object models.
Here several papers on aspects of the problem for 3D polygonal
models are described at a high level.

\subsection*{Levels of Detail}
A consensus may be emerging in favor of the representation
of complex models as a single, hierarchical data structure
that represents many levels of detail simultaneously,
from the simplified root to the fully detailed leaves.
The hierarchy is known under various names:
vertex tree~\cite{le-vdsape-97},
merge tree~\cite{xv-dvdspm-96},
vertex hierarchy~\cite{h-vdrpm-97},
progressive mesh~\cite{h-pm-96},
progressive simplicial complex~\cite{ph-psc-97}.
We will use {\em vertex tree\/} here to refer to the
generic concept.
Typically the tree is binary, with each node represting a
vertex %$v$ 
of some simplification of the original model.
The two children vertices $v_1$ and $v_2$ of their parent
$v$ are merged (or identified, or unified, or contracted),
moving up the tree to produce $v$,
which inherits all the triangles incident to its children.
Viewing the same process in reverse,
the parent $v$ splits to generate its children
at an increased level of detail.
How the coordinates of $v$ relate to those of its children
depends on the particular hierarchy implementation:
e.g., whichever of $\{ v_1, v_2, \frac{1}{2}(v_1+v_2) \}$ is 
``best''~\cite{ph-psc-97},
or a position that minimizes some geometric error~\cite{gh-ssqem-97}.
Some schemes~\cite{h-pm-96,xv-dvdspm-96,h-vdrpm-97} 
restrict $v_1$ and $v_2$ to be
connected by an edge at their model level, 
in which case the upward tree movement is an {\em edge contraction}.
This has the advantage of preserving the abstract topology\footnote{
	The reason for the qualification is that guaranteeing
	preservation of geometric simplicity requires careful
	placement of $v$, care not usually exercised
	for pragmatic reasons.  An exception is~\cite{cmo-spmusm-97}.
}
of the model,
an advantage that becomes an impediment to massive simplification of
complex models.
Thus much recent work~\cite{gh-ssqem-97,le-vdsape-97,ph-psc-97} countenances
arbitrary vertex pair identification, which may, for example,
merge separate topological components.
Before addressing which pairs of vertices should be unified,
we turn to how a model vertex tree can be utilized by a graphics system.

\subsection*{View-dependent Simplification}
The vertex tree is constructed from the fully detailed original
model in a preprocessing phase that can take anywhere from a
few seconds to (in one cited instance~\cite{ph-psc-97}) over $22$ hours.
In any one frame, the model is rendered from a list of
{\em active vertices\/}~\cite{h-vdrpm-97} which represent a variable-detail
frontier within the vertex tree~\cite{le-vdsape-97}.
Each vertex node points to relevant incident faces (triangles)
with enough information for rendering.
Before a new frame is rendered, the list of active vertices is
traversed, and a decision made whether to split a node to
increase detail, merge two nodes to simplify, or leave as is.
This decision is based on a {\em screen-space\/} error criterion.
The idea is that what matters is what the user sees---{\em object-space\/} 
geometric errors are less relevant.
Projected surface deviation~\cite{h-vdrpm-97} and silhouette preservation~\cite{le-vdsape-97}
among other heuristics have been used.
It is of course crucial that these computations be fast, and
indeed impressive real-time behavior on graphics workstations
has been achieved.

\subsection*{Construction of Hierarchy}
In contrast to reliance on screen-space error to decide which nodes
to display, geometric and topological considerations dominate the
initial tree construction.
One method, explored in~\cite{cvmtwabw-se-96}, computes inner and outer
{\em simplification envelopes\/} for an object, both within
$\epsilon$ of the original surface,
and threads a simplification between.
An algorithm for this difficult subproblem
approximates a global optimum by
greedily accumulating threading triangles that ``cover'' (in projection)
many vertices.
Successively larger values of $\epsilon$ lead to vertex clustering
that could be represented in a vertex tree.\footnote{
	The work in~\cite{cvmtwabw-se-96} is not, however, 
	geared toward constructing a hierarchy.
}

Other methods more directly follow the vertex tree structure,
choosing to unify the pair of vertices that minimize 
error~\cite{gh-ssqem-97,ph-psc-97}.
Because it is prohibitive to consider all pairs of vertices,
a crude winnowing is performed, based on topology
(vertices connected by an edge) and geometric proximity
(Euclidean distance between vertices~\cite{gh-ssqem-97};
``tight'' octree clustering~\cite{le-vdsape-97};
Delaunay edges between components~\cite{ph-psc-97}).
Some schemes assume manifold 
objects~\cite{cvmtwabw-se-96,h-pm-96,xv-dvdspm-96}; 
others permit arbitrary simplicial 
complexes~\cite{gh-ssqem-97,le-vdsape-97,ph-psc-97}.
In the surviving list of candidate pairs, a merging ``cost''
is computed for each pair, based on various geometric-based
heuristics:
estimation of the deviation of $v$ from neighboring face planes 
%to $v_1$ and $v_2$ 
by an ``error quadric'' in~\cite{gh-ssqem-97},
and a mixture of distance, area-stretching, and folding penalties
in~\cite{ph-psc-97}.
For each merge,
between-level dependencies must be carefully arranged to permit
subsequent swift navigation through the hierarchy.

There are considerable tradeoffs between speed of construction
and the ``quality'' of the resulting hierarchy,
and much of the research focus has been on the details
glossed here, which determine these tradeoffs.
We now sketch the error quadrics of~\cite{gh-ssqem-97} to
give a sense of some of these details.

\subsection*{Error Quadrics}
Define the distance $d(v,P)$ 
of a point $v = [x \, y \, z \, 1]^T$ to a set $P$ of planes
to be the sum of the squares of the distances from $v$ to
the planes in $P$. For a single plane $p = [a \, b \, c \, d]^T \in P$,
the square distance is $(v^T p) (p^T v) = v^T (p p^T) v$.
Summing this expression over all $p \in P$ yields
$d(v,P) = v^T Q v$,
where $Q$ is a $4 \times 4$ matrix.

Note that the distance between $v$ and all the planes containing
faces incident to $v$ is zero, because $v$ lies on each of these
planes.  Now consider $v$ to be the parent of two vertices
$v_1$ and $v_2$, whose unification somehow yielded $v$.
We define the cost of the merge, or the error associated with $v$,
to be $d(v,P_1)+d(v,P_2) = v^T (Q_1 + Q_2) v$, where $P_i$ are
the planes containing
faces incident to $v_i$, and $Q_i$ the corresponding matrices, $i=1,2$.
Note that if a face is incident to both $v_1$ and $v_2$, its plane will
be in both $P_1$ and $P_2$, and will be doubly weighted by
the matrix sum $Q_1 + Q_2$.

This leaves how to chose $v$.  For $Q$ fixed, 
$v^T Q v = v^T (Q_1 + Q_2) v$
is a quadric surface.  Its level surface $v^T Q v = \epsilon$ 
is a (potentially degenerate) 
ellipsoid specifying a region of space within which any $v$
has an error of at most $\epsilon$.
The $v$ used in~\cite{gh-ssqem-97} is the center of this ellipsoid.

\subsection*{Evaluation}
Currently the efficacy of a proposed simplification algorithm is 
evaluated by a mixture of run-time data, intuition, and visual appeal.
In some sense the ultimate arbiter is the (inevitable) SIGGRAPH
video.
It would seem useful to place evaluation on a more firm theoretical footing.
With much of the code in the public domain,
experimental comparisons are now appearing~\cite{cms-cmsa-98}.

\vspace{3mm}
\footnotesize
\noindent
{\bf Acknowledgements.}
I thank 
Michael Garland,
Paul Heckbert,
Hugues Hoppe,
David Luebke,
and
Dinesh Manocha
for comments.
\vspace{-1mm}
\begin{flushright}
(Written 6 March 1998)
\end{flushright}
\vspace{-3mm}
%\bibliography{33}

\begin{thebibliography}{CVM{\etalchar{+}}96}

\bibitem[CMO97]{cmo-spmusm-97}
J.~Cohen, D.~Manocha, and M.~Olano.
\newblock Simplifying polygonal models using successive mappings.
\newblock In {\em Proc. IEEE Visualization '97}, pages 395--402, 1997.

\bibitem[CMS]{cms-cmsa-98}
P.~Cignoni, C.~Montani, and R.~Scopigno.
\newblock A comparison of mesh simplification algorithms.
\newblock In {\em Computers \& Graphics}, volume~22. Pergamon Press.
\newblock To appear; {\tt http://miles.cnuce.cnr.it/cg/bibliography.html}.

\bibitem[CVM{\etalchar{+}}96]{cvmtwabw-se-96}
J.~Cohen, A.~Varshney, D.~Manocha, G.~Turk, H.~Weber, P.~Agarwal, F.~Brooks,
  and W.~Wright.
\newblock Simplification envelopes.
\newblock In {\em Proc. SIGGRAPH '96}, pages 119--128, 1996.

\bibitem[GH97]{gh-ssqem-97}
M.~Garland and P.~S. Heckbert.
\newblock Surface simplification using quadric error metrics.
\newblock In {\em Proc. SIGGRAPH '97}, pages 209--216, 1997.

\bibitem[Hop96]{h-pm-96}
H.~Hoppe.
\newblock Progressive meshes.
\newblock In {\em Proc. SIGGRAPH '96}, pages 99--108, 1996.

\bibitem[Hop97]{h-vdrpm-97}
H.~Hoppe.
\newblock View-dependent refinement of progressive meshes.
\newblock In {\em Proc. SIGGRAPH '97}, pages 189--198, 1997.

\bibitem[LE97]{le-vdsape-97}
D.~Luebke and C.~Erikson.
\newblock View-dependent simplification of arbitrary polygonal environments.
\newblock In {\em Proc. SIGGRAPH '97}, pages 199--208, 1997.

\bibitem[PH97]{ph-psc-97}
J.~Popovi{\'c} and H.~Hoppe.
\newblock Progressive simplicial complexes.
\newblock In {\em Proc. SIGGRAPH '97}, pages 217--224, 1997.

\bibitem[XV96]{xv-dvdspm-96}
J.~Xia and A.~Varshney.
\newblock Dynamic view-dependent simplification for polygonal models.
\newblock In {\em Proc. IEEE Visualization '96}, pages 327--334, 1996.

\end{thebibliography}
\newcommand{\etalchar}[1]{$^{#1}$}

\end{document}